\documentclass[graybox]{svmult}
\usepackage{type1cm}        
\usepackage{makeidx}         
\usepackage{graphicx}        
\usepackage{multicol}        
\usepackage[bottom]{footmisc}

\usepackage{newtxtext}       %
\usepackage[varvw]{newtxmath}       

\usepackage{graphicx}
\usepackage{tikz}
\usepackage{relsize}
\usetikzlibrary{arrows,calc,chains,positioning,shapes.geometric,fit}
            
\tikzset{fontscale/.style = {font=\relsize{#1}}}                
\tikzstyle{sensor} = [rectangle, rounded corners, minimum width=0.5cm, minimum height=1cm,
    text width=1.5cm, text centered, draw=black, fill=blue!30]

\tikzstyle{au} = [rectangle, rounded corners, minimum width=0.5cm, minimum height=1cm,
    text width=1.5cm, text centered, draw=black, fill=red!30]

\tikzstyle{gadget} = [rectangle, rounded corners, minimum width=0.5cm, minimum height=1cm,
    text width=1.5cm, text centered, draw=black, fill=green!30]

\tikzstyle{service} = [rectangle, rounded corners, minimum width=0.5cm, minimum height=1cm,
    text width=1.5cm, text centered, draw=black, fill=orange!30]

\tikzstyle{requiredstep} = [rectangle, minimum width=0.5cm, minimum height=1cm,
    text width=2.5cm, text centered, draw=black, fill=white]

\tikzstyle{optionalstep} = [rectangle, minimum width=0.5cm, minimum height=1cm,
    text width=2.5cm, text centered, draw=black, fill=black!10]

\tikzstyle{database} = [cylinder,
cylinder uses custom fill,
cylinder body fill=yellow!50,
cylinder end fill=yellow!50,
shape border rotate=90,
aspect=0.25,
minimum width=0.5cm, minimum height=1cm, text width=1.5cm, text centered,
draw]
\tikzstyle{line} = [draw, -latex]

\usepackage{eso-pic}
\newcommand\AtPageUpperMycenter[1]{\AtPageUpperLeft{%
 \put(\LenToUnit{0.12\paperwidth},\LenToUnit{-1cm}){%
     \parbox{1.2\textwidth}{\raggedleft\fontsize{9}{11}\selectfont #1}}%
 }}%
\newcommand{\conf}[1]{%
\AddToShipoutPictureBG*{%
\AtPageUpperMycenter{#1}
}
}

\newcommand{\system}{DataX}

\makeindex             

\begin{document}

\title*{\system: A system for Data eXchange and transformation of streams}

\author{Giuseppe Coviello \and
Kunal Rao \and
Murugan Sankaradas \and
Srimat Chakradhar}

\institute{Giuseppe Coviello \and Kunal Rao \and Murugan Sankaradas \and Srimat Chakradhar \at NEC Laboratories America, Inc., Princeton, NJ 08540, \email{\{giuseppe.coviello,kunal,murugs,chak\}@nec-labs.com}}



\maketitle
\conf{The 14th International Symposium on Intelligent Distributed Computing, Sep 16-18, 2021, Italy}

 \abstract{The exponential growth in smart sensors and rapid progress in 5G networks is creating a world awash with data streams. However, a key barrier to building performant multi-sensor, distributed stream processing applications is high programming complexity. We propose \system, a novel platform that improves programmer productivity by enabling easy exchange, transformations, and fusion of data streams. \system\ abstraction simplifies the application's specification and exposes parallelism and dependencies among the application functions (microservices). \system\ runtime automatically sets up appropriate data communication mechanisms, enables effortless reuse of microservices and data streams across applications, and leverages serverless computing to transform, fuse, and auto-scale microservices. \system\ makes it easy to write, deploy and reliably operate distributed applications at scale. Synthesizing these capabilities into a single platform is substantially more transformative than any available stream processing system.}



%
%
\section{Introduction}
\label{introduction}
Smart sensors today sense, digitize and communicate over the internet~\cite{FortinoTrunfio2014}. The exponential growth in such sensors~\cite{iot-link}, and progress in 5G networks~\cite{ericsson-link}, is creating a world awash with digital streams. In the future, this exponential growth in smart sensors will result in billions of digital data streams, each describing an increasingly smaller aspect of the physical or digital worlds in greater detail. A rich understanding of these complex worlds, which will be impossible to do by using information from any single sensor, will inevitably require the fusion of information in various data streams.

A key barrier to building performant multi-sensor distributed applications is high programming complexity, which results in the low productivity of application developers. 
\textit{First}, they have to express the functionality of an application as a collection of interacting microservices~\cite{microservices-link,microservices} that form an efficient processing pipeline. There is no standard way to do this for arbitrary applications. \textit{Second}, they have to design and specify appropriate data communication mechanisms among the distributed microservices, and this has a big impact on the performance of the application. \textit{Third}, they have to worry about the actual hardware on which the microservices will run and therefore, manage the underlying computing infrastructure as well. \textit{Fourth}, if any of the microservices need to maintain state, then developers are forced to manage appropriate databases.

Scaling multi-sensor stream processing applications, and ensuring reliable operation in the face of software and hardware failures, is yet another problem that application developers have to address. Furthermore, if the developers want to extend their application pipeline by reusing microservices from another application pipeline, then there is no easy way to do so, other than obtaining the microservice, understanding the interfaces, implementing appropriate data communication, and then embedding the microservice within the desired application pipeline. This is very tedious and cumbersome. 
Removing such barriers and improving application developer productivity is the focus of this paper.


In this paper, we present \system, a platform that improves programmer productivity by enabling easy exchange, transformation and management of data streams in complex, multi-sensor distributed stream processing applications. We propose simple programming abstractions and SDKs to specify multi-sensor applications. Developers define and register objects like sensors, drivers, streams, analytics units, actuators, and gadgets, all of which enable succint specification of the overall application pipeline (Section~\ref{abstraction}). 

\system\ runtime (Sections~\ref{implementation} and \ref{application}) automatically determines appropriate data communication mechanisms among the application's objects, including network connections, serialization and deserialization of data streams. \system\ runtime provides serverless computing, and the developers only provide the business logic for different types of analytics processing. Application-specific allocation, scheduling and execution on the underlying distributed computing resources, as well as auto-scaling and reliable operation, is automatically managed by \system. For stateful applications, \system\ manages and exposes a database, which can readily be used by the developer. \system\ also makes it easy for developers to effortlessly reuse any registered microservice in any application pipeline by just subscribing to the service's output streams. 

The synthesis of all of these capabilities into a single platform like DataX is substantially more transformative than any known stream processing system. To the best of our knowledge, there is no other system that addresses all these developer concerns and simplifies multi-sensor application development, deployment and reliable operation on different distributed computing fabrics.

\section{\system\ abstraction}
\label{abstraction}
\system\ offers a simple abstraction for building complex, multi-sensors applications. It allows developers to focus on the business logic rather than the complex intricacies of writing, deploying, or operating containerized applications on distributed systems. 
Application developers use \system\ abstractions like  \textit{sensors}, \textit{gadgets}, \textit{streams}, \textit{analytics units} and \textit{actuators} to build distributed stream processing applications. 

A \textit{sensor} is a device that produces raw data. Examples of sensors include cameras, location sensors, temperature and pressure sensors, LIDAR, and radar. 
Applications process and analyze raw data from sensors to generate insights.
Sensors may have wired or wireless networking capability, or they may be physically attached to a computing device through interfaces like USB. Sensors are the starting point of a \system\ application.

A \textit{gadget} is a (physical or virtual) device that can be controlled by using insights derived from data analysis. 
Examples of gadgets include entry/exit gates, displays, LED arrays, or graphical user interfaces. 
Like sensors, some gadgets may have networking capability, while others may be attached to computing devices.
Gadgets are ideally the ending point of a \system\ application. Not all \system\ applications may have gadgets. Also, a physical or virtual device can be both a sensor and a gadget if the device produces data, and the device can also be controlled. For example, a camera can produce a video stream; and it also allows remote management to configure camera parameters. 


A data \textit{stream} is a continuous flow of homogeneous discrete messages. Some streams have only data produced by the sensors, while others can contain the insights obtained by analyzing and fusing the data from sensors. An \textit{analytics unit} (AU) processes on one or more input streams to mine insights (i.e., analytical processing), or transforms the input streams for future analysis, and produces output streams with rich insights obtained from analytical processing. 

For example, \system\  produces a stream containing the frames (images) captured for each registered camera. An object detector, which is an example of an analytics unit, can subscribe to the stream, process the stream, and generate an output \textit{augmented} stream that has all the objects detected in each frame. The augmented stream can be used as input to other analytics units or \textit{actuators}, which control various gadgets by using the information in the input streams. 


\begin{figure}[b]
    \centering
    \includegraphics[width=0.78\textwidth]{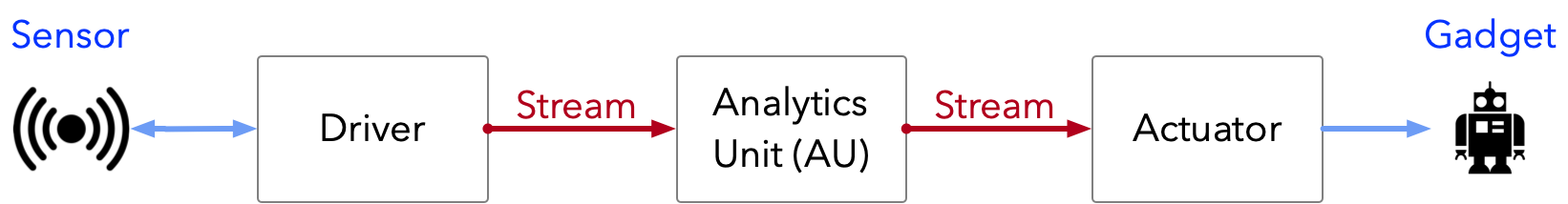}
    \vskip -0.05in
    \caption{A generic \system\ application}
    \label{example-application}
\end{figure}

Figure~\ref{example-application} shows the analytics processing pipeline in a generic \system\ application. Application developer provides the business logic (or full docker image) for generating a stream from a sensor (i.e., a driver), analyzing a stream or fusing multiple streams (i.e., an analytics unit), or controlling a gadget (i.e., actuator). 
Application developers can also re-use business logic functions already available in \system\ like drivers for cameras, actuators for entry/exit turnstiles, and analytics units for recognizing faces or detecting objects. \system\ also provides a simple SDK (software development kit) for writing drivers, analytics units, and actuators. 
\system\ supports \textit{serverless} computing. Application developers only provide business logic to implement drivers, AUs, and actuators. For stateful applications, \system\ provides an abstraction for creating databases and attaching them to drivers, analytics units, and actuators. \system\ installs and maintains the databases, while applications are responsible for the content in the databases (like creating a schema or storing data). 

%



\section{Benefits}
\textbf{Simple abstraction.}
\system\ abstractions and SDK are intuitive and easy to use. Analytics applications typically receive input data from one or more sensors using drivers. The sensor data from drivers then undergoes a series of transformations and processing by analytics units, which produce output streams that can activate a gadget through an actuator. Sensors, drivers, analytics units, actuators, and gadgets are all simple \system\ abstractions,
which make it easier to write, as well as deploy and reliably operate distributed applications at scale. \\
\\
\textbf{Automated data communication}
\system\ automatically chooses appropriate communication mechanisms to exchange data among drivers, analytics units, and actuators to accomplish application-specific goals.
In addition, \system\ handles the network connections as well as manages serialization and deserialization of data when data is being transferred among drivers, analytics units and actuators. By handling all aspects of data communication, \system\ relieves application developers from additional effort to design application-specific data communication.\\
\\
\textbf{Serverless stream processing.}
In addition to writing the software components, application developers typically also have to worry about the computing resources necessary to execute these components. This is often not so straightforward in a distributed system, where multiple applications are running simultaneously on a shared computing infrastructure that is dynamic and heterogeneous. This daunting task of specifying and managing execution on the underlying compute resources is automatically handled by \system
. This is similar to serverless computing, where developers only provide the business logic 
and actual execution is handled transparently.\\
\\
\textbf{Easy state management.}
AUs sometimes need to manage state to provide specific functionality. 
Developers usually have to handle state by themselves, and there is little to no support from the underlying platform. \system\ makes this state management easy by exposing in-built database management systems and the associated databases. Developers can choose the specific database, create the desired schema, and manage the desired content/state. This \system\ platform-provided state management makes it easy for developers to manage any state within and across AUs.\\
\\
\textbf{Effortless data streams reuse.}
One of the most beautiful features about \system\ is that it allows developers to reuse data streams produced by third-party applications effortlessly. This fosters incremental development and 
there is no need to reinvent the wheel; developers already know what sensors are registered with \system\ and what kind of augmented streams are produced from various sensor data streams. Using this knowledge from \system, developers can quickly build new and more exciting sensor-fusion applications by reusing existing raw or augmented data streams.

\section{Design and implementation}
\label{implementation}
The target of \system\ is programmers, and the focus is on programming abstractions for application developers. These abstractions simplify application development, and make distributed stream processing applications easier to write. Furthermore, data scientists who are users of \system\ are likely not familiar with Kubernetes or other container management systems. Enabling scientists to rapidly get their work done, and focus on their domain-specific streaming applications rather than distributed systems, is the primary goal of \system.

Accordingly, the implementation of \system\ ensures that developers and programmers who use \system\ platform are largely unaware that they are running on top of Kubernetes, the de-facto standard for cloud-native development. Kubernetes is an open-source orchestrator for deploying containerized applications. It is a complex system that makes applications faster to deploy and more reliable to operate at scale. Although Kubernetes simplifies the deployment and operations of distributed applications, it does little to make the development of distributed applications easy. With \system\ as a complete abstraction on top of Kubernetes, we can now not only deploy and reliably operate containerized applications at scale, but we can also simplify application development and make stream processing applications easier to write.

There are at least two different options for realizing \system\ on top of Kubernetes. \textit{First} option is to use stock Kubernetes API server out of the box. Applications developers use Kubernetes's resources like Pods, StatefulSets, and ReplicaSets, to describe the application's desired state, and  Kubernetes deploys the application and strives to ensure that the current state matches the declarative description of desired state~\cite{kubernetes-uar}. However, in this option, Kubernetes does not understand the workload running in Kubernetes, objects like StatefulSets or ReplicaSets, or the application's domain-specific knowledge of running a specific application. \textit{Second}, and more performant option is to extend the Kubernetes API server by adding new functionality using operators \cite{operators}, which allow encapsulation of domain-specific knowledge of running a specific stream processing application. \system\ extends Kubernetes API by defining operators, allowing \system\ to take advantage of the vast Kubernetes ecosystem of tools.

\system\ defines \textit{custom resources} as extensions of Kubernetes API that are not available in a default Kubernetes installation. A resource is an endpoint in Kubernetes API that stores a collection of API objects of a certain kind; for example, the built-in Pods resource contains a collection of Pod objects. Once a custom resource is installed, we can create and access its objects using kubectl, just as we do for built-in resources like Pods. \system\ makes components like driver, AU, actuator, sensor, gadget, stream, and databases as first-class citizens of Kubernetes by installing them as custom resources.

\system\ also extends Kubernetes API by defining a new operator, which is a software extension to Kubernetes that makes use of custom resources to manage \system\ applications and their components. The Operator pattern aims to capture the key aim of a human operator managing a service or set of services. Human operators who look after specific applications and services have deep knowledge of how the system should behave, how to deploy it, and how to react if there are problems. The Operator pattern~\cite{operators} captures how you can write code to automate a task beyond what Kubernetes itself provides. \system\ Operator adds one more custom resource to Kubernetes. The Operator provides the logic for monitoring and maintaining the resources it defines, which means that the Operator will take actions based on the resource's state defined by \system.

Extending Kubernetes API server leads to multiple benefits. \system\ administrators can interact with Kubernetes using familiar and standard interface (YAML resource descriptions, kubectl, etc.). Also, \system\ components like driver, AU, actuator, sensor, gadget, stream, and database become Kubernetes-managed objects, just like a pod, for example. This way,  \system\ does not have to implement new, custom software modules to auto-scale, store or update application domain-specific components.

\begin{figure}[b]
    \centering
    \includegraphics[width=0.71\textwidth]{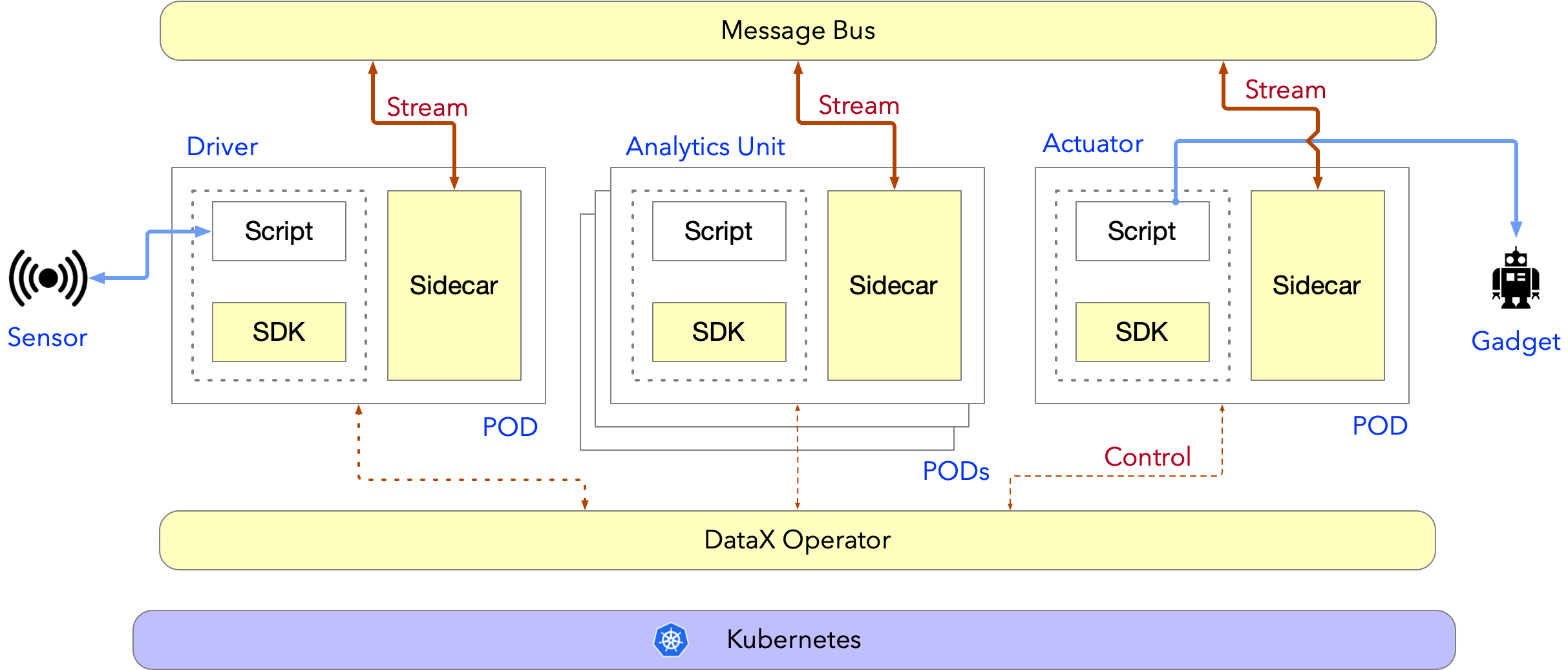}
    \vskip -0.1in
    \caption{\system\ Components}
    \label{components}
\end{figure}

Figure~\ref{components} shows (in yellow) the components of \system: a Kubernetes Operator defines and manages \system's custom resources, a message bus facilitates communication among the microservices of the application, a sidecar container in each Pod handles data communication among various \system\ entities (drivers, AUs or actuators), and language-specific SDKs enable easy use of \system\ APIs.\\
\\
\textbf{\system\ Operator.}
The \system\ Operator extends Kubernetes to support the entities defined in Section~\ref{abstraction}; it provides the CRDs (custom resource definition) and logic for the driver, AU, actuator, sensor, gadget, and stream.
The \system\ Operator takes necessary actions to ensure that all \system\ applications are in a coherent state at all times. It also protects the system from user's actions that might bring the system into an unrecoverable incoherent state. For example, uninstalling a driver while a sensor is being used can bring the system into an incoherent state, which will be detected and avoided.\\
\\
\textbf{Drivers, AUs, and actuators.}
To register a driver, an AU, or an actuator, the user provides the name, either a script (pure serverless) or a docker image containing the business logic and (optionally) the configuration schema. When the user provides just a script, \system\ will automatically create a docker image for executing the script. When the user requests an upgrade of drivers, AUs, and actuators, \system\ Operator will automatically cascade the upgrade to running instances. \system\ Operator ensures the coherency of the upgrade by enforcing that new configuration schemas are compatible with the schemas of the running instances. When performing an upgrade, optionally, the user can provide a script to convert the configuration schemas; in such case, \system\ Operator will accept the upgrade only if the script can be executed successfully for all the running instances (compatible configurations are generated for all instances). Also, if a user requests the deletion of a driver, AU, or actuator, then \system\ Operator will check if the entity is currently in use and refuse the operation if there is already a running instance for that entity.\\
\\
\textbf{Sensors, streams, and gadgets.}
When registering a sensor, \system\ Operator ensures that (a) the required driver is installed, and (b) the driver configuration schema provided by the user is compatible with the configuration schema expected by the installed driver. \system\ Operator will also maintain the driver's running instance (on appropriate computing resources) as long the sensor is registered. For example, if the sensor is physically attached to a computing node through a USB interface, then \system\ Operator will maintain a running instance on the same computing node. A registered sensor always generates an output stream that has the same name as the sensor. The user requests to create augmented streams by providing: (a) AU that generates the stream, (b) the input streams, and (c) the AU configuration. \system\ Operator checks that the AU is available, the configuration is compatible, and the input streams are registered; \system\ Operator, unless the user requests a fixed number of instances, auto-scales the number of instances of the AU. \system\ Operator performs similar operations when the user requests to register a new gadget. 
Before deleting any sensors or streams, \system\ Operator ensures that they are not input to produce other streams.\\
\\
\textbf{Message bus.}
Communication among microservices in a distributed application is typically accomplished by using either (some variant of) REST/HTTP or a messages bus/distributed queue \cite{bus-microservices}. \system\ uses NATS, which is a lightweight, distributed, and scalable message queue~\cite{nats}, for automatically setting up communication mechanisms among various entities like drivers, AUs, and actuators. \system\ Operator manages the deployment and configuration of NATS (the \system\ Operator uses Kubernetes DaemonSet construct to deploy a NATS cluster). The NATS cluster uses authentication and authorization, and only services deployed on \system\ will be able to connect to the NATS server (they will be able to subscribe and publish only on the defined and registered streams).\\
\\
\textbf{\system\ Sidecar.}
\system\ Sidecar is a containerized application that \system\ Operator runs alongside each instance of a user-provided driver, AU, or actuator (in the same Pod). The main role of the \system\ Sidecar is to automatically manage data communication (it manages the connection, subscriptions, and publishing to the messages bus). Also, \system\ Sidecar monitors the health of the user's application; it exposes, using REST API, the metrics such as the systems resources utilization and the number of messages received, dropped, and published. \system\ Operator and Kubernetes use those metrics to ensure that all the components are working correctly, and these metrics also drive the auto-scaling process.\\
\\
\textbf{\system\ SDKs.}
\system\ provides SDKs for several programming languages (C, Go, JavaScript, Python). SDKs provide APIs for retrieving configuration, receiving data, and publishing data. These APIs are implemented using the idiom of the programming language; for example, SDK for Python exposes a class \texttt{\system} having three public methods: \texttt{get\_configuration()}, to retrieve the configuration in the form of a dictionary (pairs of key-values); \texttt{next()} to obtain a message from one of the input streams, this method returns a tuple of two values, the name of the stream and the message in the form of a dictionary where keys are strings; \texttt{emit(message)} for publishing a new message, which is a dictionary where keys are strings. \system\ SDKs are simple and lightweight. \system\ Sidecar does most of the work in managing data communication, and the SDKs provide an interface between \system\ Sidecar and the business logic (SDKs and \system\ Sidecar use shared memory for communicating with each other).

\section{Distributed application development using \system}
\label{application}

\setlength\intextsep{3pt}
\begin{figure}[b]
    \centering
    \includegraphics[width=0.8\textwidth]{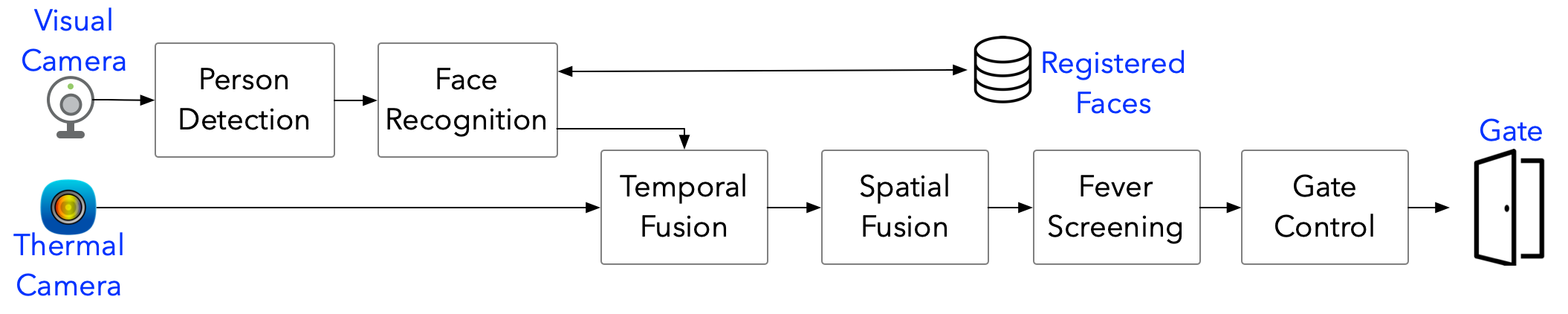}
    \vskip -0.1in
    \caption{Fever screening application on \system}
    \label{ebt-streams}
\end{figure}

We designed a popular fever screening application~\cite{EDGEDL2021} pipeline, shown in Figure~\ref{ebt-streams} on \system. 
This complex \system\ application has two sensors, two drivers, one actuator, one gadget, five analytics units, and one database. Application developer only provides the business logic for these entities, and \system\ takes care of stream data communication, serverless computing, state management, and stream reuse. Also, \system\ leverages Kubernetes to deploy and reliably operate the application at scale. The synthesis of these capabilities into a single product like \system\ is substantially more transformative than any available stream processing system.

\section{Related Work}
\label{related-work}

There are several open source and proprietary stream processing systems like Flink \cite{flink-link}, Spark Streaming \cite {spark-streaming-link}, Storm \cite{storm-link}, Heron \cite{heron-link}, ksqlDB \cite{ksqldb-link}, Esper \cite{esper-link}, Hazelcast Jet \cite{hazelcast-jet-link}, Hitachi Streaming Data Platform \cite{hitachi-streaming-data-platform}, IBM Streams \cite{ibm-streams}, Microsoft Azure Stream Analytics \cite{azure-stream-analytics}, Oracle Stream Analytics \cite{oracle-stream-analytics}, SAS Event Streaming \cite{sas-event-stream-processing}, TIBCO Streaming \cite{tibco-streaming}, etc. that provide capabilities to handle data streams. 
However, these systems have full-fledged and specific programming models and supporting runtime systems, which restricts the way applications are developed in order to be deployed on these systems. In contrast, \system\ provides flexibility in choice of programming language and simplifies programming using simple abstractions.

There are systems for stream data integration like Apache Airflow \cite{apache-airflow}, Cloudera DataFlow \cite{cloudera-dataflow}, Data Stream Manager \cite{data-stream-manager}, IBM DataStage \cite{ibm-datastage}, Informatica Data Engineering Streaming \cite{data-engineering-streaming}, StreamSets Data Collector \cite{streamsets-data-collector}, Striim Platform \cite{striim-platform}, etc. that deal with ingestion and storage of data streams. Such storage of data streams is not included in \system\ to avoid additional storage overhead.


AWS Lambda \cite{aws-lambda} is a service provided by Amazon for running serverless computing, wherein the developer provides just a ``Lambda" function that operates on data. These functions can be combined and integrated with Amazon Kinesis \cite{kinesis-link} to create a pipeline for processing streaming data in real-time \cite{aws-kinesis-lambda} \cite{aws-kinesis-lambda-article}. These services i.e. Kinesis and Lambda operate as separate services. 
\system, on the other hand, combines functionality of Amazon Kinesis and AWS Lambda into a middleware product that can be deployed anywhere, including resource constrained edge environments.

\section{Conclusion}
\label{conclusion}
A key barrier to building multi-sensor distributed stream processing applications is high programming complexity. \system\ makes it easy to write, deploy, and reliably operate distributed stream processing applications at scale. Although \system\ uses Kubernetes and NATS internally, \system\ can be implemented on other container management or message queuing systems. \system\ can be used as an independent platform or embedded in any streaming application. Also, \system\ applications can run on-premise, in the cloud, or any edge-cloud computing fabric. 
\bibliographystyle{spmpsci}

\end{document}